\documentclass[11pt]{article}
\usepackage{amsmath,amssymb, color,epsfig,cite}
\usepackage{graphicx}
\usepackage{setspace}
\usepackage{pstricks,pst-node,pst-plot, mathtools, color, epsfig}

\usepackage{mathrsfs}

\usepackage{amsfonts}
\newcommand{\hoch}[1]{$\, ^{#1}$}

\makeatletter
\@addtoreset{equation}{section}
\makeatother

\newcommand{\be}{\begin{equation}}
\newcommand{\ee}{\end{equation}}
\newcommand{\bea}{\setlength\arraycolsep{2pt} \begin{eqnarray}}
\newcommand{\eea}{\end{eqnarray}}
\newcommand{\nn}{\nonumber}
\newcommand{\half}{{\textstyle{\frac{1}{2}}}}

\def\ft#1#2{{\textstyle{\frac{\scriptstyle #1}{\scriptstyle #2} } }}
\def\fft#1#2{{\frac{#1}{#2}}}

\def\0{{\sst{(0)}}}
\def\1{{\sst{(1)}}}
\def\2{{\sst{(2)}}}
\def\3{{\sst{(3)}}}
\def\4{{\sst{(4)}}}
\def\5{{\sst{(5)}}}
\def\6{{\sst{(6)}}}
\def\7{{\sst{(7)}}}
\def\8{{\sst{(8)}}}
\def\sst#1{{\scriptscriptstyle #1}}

\def\del{{\partial}}

\def\crampest{\medmuskip = 1mu plus 1mu minus 1mu}
\def\uncramp{\medmuskip = 4mu plus 2mu minus 4mu}

\def\F{\mathcal{F}}
\def\G{\mathcal{G}}

\allowdisplaybreaks

\begin{document}

\begin{flushright}
\hfill{\hfill{MI-TH-1931}}

\end{flushright}

\vspace{15pt}
\begin{center}
{\Large {\bf Taub-NUT from the Dirac monopole}}

\vspace{15pt}
{\bf Hadi Godazgar\hoch{1}, Mahdi Godazgar\hoch{2} and 
C.N. Pope\hoch{3,4}}

\vspace{10pt}

\hoch{1} {\it Max-Planck-Institut f\"ur Gravitationsphysik (Albert-Einstein-Institut), \\
M\"uhlenberg 1, D-14476 Potsdam, Germany.}

\vspace{10pt}

\hoch{2} {\it School of Mathematical Sciences,
Queen Mary University of London, \\
Mile End Road, E1 4NS, United Kingdom.}

\vspace{10pt}

\hoch{3} {\it George P. \& Cynthia Woods Mitchell  Institute
for Fundamental Physics and Astronomy,\\
Texas A\&M University, College Station, TX 77843, USA.}

\vspace{10pt}

\hoch{4}{\it DAMTP, Centre for Mathematical Sciences,\\
 Cambridge University, Wilberforce Road, Cambridge CB3 OWA, UK.}

 \vspace{15pt}

 August 16, 2019

\vspace{20pt}

\underline{ABSTRACT}
\end{center}

\noindent

Writing the metric of an asymptotically flat spacetime in Bondi coordinates
provides an elegant way of formulating the Einstein equation as a
characteristic value problem.  In this setting, we find that
a specific class of asymptotically flat spacetimes, including stationary 
solutions, contains a Maxwell gauge field as free data.  Choosing this 
gauge field to correspond to the Dirac monopole, we derive the 
Taub-NUT solution in Bondi coordinates.

\noindent

\thispagestyle{empty}

\vfill
E-mails: hadi.godazgar@aei.mpg.de, m.godazgar@qmul.ac.uk, pope@physics.tamu.edu

\pagebreak

\section{Introduction}

Asymptotically flat spacetimes have been extensively studied over the last 
50 years, particularly, recently, in the context of gravitational wave 
detections \cite{Bishop2016} and asymptotic  symmetry 
groups \cite{Strom:lec,Compere:2018aar}.  In this paper, motivated by 
recent work on dual gravitational charges 
\cite{dual0, dualex, porrati, softnuts}, we show how treating asymptotically 
flat spacetimes in terms of a characteristic value problem \cite{Winicour2009} 
provides an intriguing way of viewing the Dirac magnetic monopole
as a progenitor of the Taub-NUT spacetime.\footnote{The relation between 
the Dirac monopole and Taub-NUT solutions is also 
encountered in a different setting: that of the double copy 
\cite{Monteiro:2014cda, Luna:2015paa}, where, the focus of the investigation 
is on the (double) Kerr-Schild ansatz and the key insight is that the 
Kerr-Schild null vector(s) may be related to a Maxwell field.}

  Our starting point is to consider a general class of asymptotically flat
metrics, written in a Bondi coordinate system $(u,r,x^I=\{\theta,\phi\})$, 
such that the metric takes the form\footnote{See section 2 of Ref.\ \cite{fakenews} for a more in-depth discussion of asymptotically-flat spacetimes and the notation we will use in this paper.}
\begin{equation} \label{AF}
 d s^2 = - F e^{2 \beta} du^2 - 2 e^{2 \beta} du dr + 
r^2 h_{IJ} \, (dx^I - C^I du) (dx^J - C^J du),
\end{equation}
with the metric functions satisfying the following fall-off conditions at 
large $r$:
\begin{align}
 F(u,r,x^I) &= 1 + \frac{F_0(u,x^I)}{r} + \frac{F_1(u,x^I)}{r^2} + \frac{F_2(u,x^I)}{r^3} + \frac{F_3(u,x^I)}{r^4} + o(r^{-4}), \notag \\[2mm]
 \beta(u,r,x^I) &= \frac{\beta_0(u,x^I)}{r^2} + \frac{\beta_1(u,x^I)}{r^3} + \frac{\beta_2(u,x^I)}{r^4} + o(r^{-4}), \notag \\[2mm] 
 C^I(u,r,x^I) &= \frac{C_0^I(u,x^I)}{r^2} + \frac{C_1^I(u,x^I)}{r^3} + \frac{C_2^I(u,x^I)}{r^4} + \frac{C_3^I(u,x^I)}{r^5} + o(r^{-5}), \notag \\[2mm] \label{met:falloff}
 h_{IJ}(u,r,x^I) &= \omega_{IJ} + \frac{C_{IJ}(u,x^I)}{r} + \frac{C^2 \omega_{IJ}}{4 r^2} + \frac{D_{IJ}(u,x^I)}{r^3} + \frac{E_{IJ}(u,x^I)}{r^4} + o(r^{-4}),
\end{align}
where $\omega_{IJ}$ is the standard metric on the round 2-sphere with 
coordinates $x^I=\{\theta, \phi\}$ and  $C^2 \equiv C_{IJ} C^{IJ}$.  
Moreover, a residual gauge freedom allows us to require that
\begin{equation} \label{det:h}
 h =\omega,
\end{equation}
where $h \equiv \textup{det}(h_{IJ})$, and $\omega 
                      \equiv \textup{det}(\omega_{IJ}) =\sin\theta$.  
Following Ref.\ \cite{softnuts}, we do not impose on the fields
defined above any regularity conditions on the 2-sphere.

   One may introduce a complex null frame of vector fields 
$e_\mu{}^a=(\ell^a,n^a,m^a,\bar{m}^a)$ with inverse $E^\mu{}_a$,
\begin{equation}
 g_{ab} = E^\mu{}_a E^\nu{}_b \ \eta_{\mu \nu}, 
\qquad \eta_{\mu \nu} = \text {{\footnotesize $ \begin{pmatrix}
    \begin{matrix} 0 & -1 \\ -1 & 0 \end{matrix} & \mathbf{0} \\
        \mathbf{0} & \begin{matrix} 0 & 1 \\ 1 & 0 \end{matrix}
\end{pmatrix}$ }},
\end{equation}
where
\begin{align}
 \ell &= \frac{\partial}{\partial r}, \qquad n =  e^{- 2 \beta} \Bigg[ \frac{\partial}{\partial u} - \half F \frac{\partial}{\partial r} + C^I \frac{\partial}{\partial x^I} \Bigg], \qquad m = \frac{\hat{m}^I}{r}  \frac{\partial}{\partial x^I}, \notag\\
 \ell^\flat &= - e^{2\beta} du, \qquad n^{\flat} = - \Big( dr + \frac{1}{2} F du \Big), \qquad m^{\flat} = r\, \hat{m}_I\, (dx^I - C^I du),
 \label{AF:frame}
\end{align}
and 
\begin{equation}
 2 \hat{m}^{(I} \bar{\hat{m}}^{J)} = h^{IJ},
\end{equation}
with $h^{IJ}$ the matrix inverse of $h _{IJ}$.

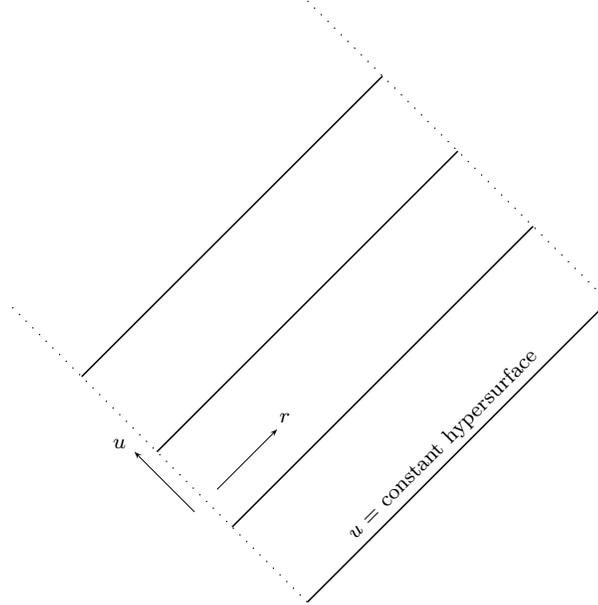
\begin{figure}[h!] \label{fig}
\begin{center}
\begin{pspicture}(0,0)(-1,7.5)
\psline[linewidth=0.2mm, linestyle=dotted](0,8)(4,4)
\psline[linewidth=0.2mm, linestyle=dotted](0,0)(-4,4)
\psline[linewidth=0.2mm, linestyle=solid](4,4)(0,0)
\psline[linewidth=0.2mm, linestyle=solid](3,5)(-1,1)
\psline[linewidth=0.2mm, linestyle=solid](2,6)(-2,2)
\psline[linewidth=0.2mm, linestyle=solid](1,7)(-3,3)
\rput{45}(1.8,2.1){\scriptsize{$u=$ constant hypersurface}}
\psline[linewidth=0.1mm, linestyle=solid]{->}(-1.5,1.2)(-2.3,2.0)
\rput(-2.5,2.1){\scriptsize{$u$}}
\psline[linewidth=0.1mm, linestyle=solid]{->}(-1.2,1.5)(-0.4,2.3)
\rput(-0.3,2.45){\scriptsize{$r$}}
\end{pspicture}
\end{center}
\caption{Hypersurfaces defining a characteristic value problem.}
\end{figure}

Given the choice of Bondi coordinates, the formalism most adapted to the 
problem of constructing solutions from initial data is the characteristic 
approach \cite{Winicour2009}.  In the characteristic value problem, 
such as that defined by Bondi coordinates, the spacetime is viewed as a 
foliation of null hypersurfaces, called the characteristic surfaces,
corresponding to the level sets of $u$ 
(see figure \ref{fig}).  

The vacuum\footnote{One could, of course, include matter with corresponding energy-momentum tensor satisfying appropriate fall-off conditions.  However, in this paper, we simply consider the vacuum Einstein equation.} Einstein equation may then be divided into three types of equations \cite{Winicour2009}:
\begin{itemize}
 \item[{\bf 1.}] {\bf Hypersurface equations}: 
these are equations that hold in 
each $u=$\, constant hypersurface and are of the form 
 \begin{equation}
  \partial_r \F = H_\F(\F,\G), \label{hseqn}
 \end{equation}
 where $\F$ denotes the set of hypersurface variables, which in Bondi 
coordinates correspond to $\{\beta, F, C^I\}$, while $\G$ denotes evolution 
variables, which in Bondi coordinates corresponds to $h_{IJ}.$ The 
operator $H_\F$, as well as $H_\G$ defined below, is
non-linear in derivatives with respect to the three
hypersurface coordinates.  The structure of the hypersurface equations
is such that the $r$-dependence of the right hand side is always 
explicit, and therefore the hypersurface variables may simply be
determined by integrating with respect to $r$.

 \item[{\bf 2.}]  {\bf Evolution equations}: these are of the form
 \begin{equation}
  \partial_u \partial_r \G = H_\G(\F,\G, \partial_u \G). \label{evoleqn}
 \end{equation}
 Note that there are no second order derivatives in time.

 \item[{\bf 3.}]  {\bf Conservation equations}: 
these are satisfied on $r=$ constant 
hypersurfaces transverse to the characteristics, and 
are of the form
  \begin{equation}
  \partial_u \F = h_\F(\F,\G, \partial_u \G),
 \end{equation}
where $h_\F$ is some non-linear operator in the two angular
coordinates in the $u=$ constant 
hypersurfaces. These are to be thought of as conservation equations rather
than evolution equations because their structure is
such that once they are satisfied on a particular $r=$ constant hypersurface,
they are guaranteed to hold for all values of $r$. 

\end{itemize}

In the complex null frame introduced above, these groups correspond to 
the following components of the Einstein equation:
\bea
\hbox{\bf Hypersurface equations}:&&
\ell^\mu\,\ell^\nu\, G_{\mu\nu}=0\,,\quad
\ell^\mu\,n^\nu\, G_{\mu\nu}=0\,,\quad
\ell^\mu\,m^\nu\, G_{\mu\nu}=0\,.\qquad \label{constraints}\\
\hbox{\bf Evolution equations}:&& 
\label{uderivs}
  m^\mu\, m^\nu\, G_{\mu\nu}=0\,.\\
\hbox{\bf Conservation equations}:&&
\label{cons}
  n^\mu\,n^\nu\, G_{\mu\nu}=0\,,\quad
n^\mu\,m^\nu\, G_{\mu\nu}=0\,.
\eea
Note that the $m^\mu\, \bar{m}^\nu\, G_{\mu\nu}=0$ component is 
automatically  
satisfied if we assume the other components to hold \cite{sachs}.  
Since we are assuming 
specific expansions in inverse powers of $r$ for the metric 
components,\footnote{This assumption is formally consistent \cite{bondi, sachs} in 
the sense that assuming such a fall off for the free initial data 
$h_{IJ}(u_0,r,x^I)$ implies a solution of the form defined in 
equation \eqref{met:falloff}.} at least for the first few orders, 
the corresponding hypersurface equations simplify to algebraic equations 
at fixed $u$ 
for all scalar and vector fields on the right hand side of the fall-off 
conditions for $\{\beta, F, C^I\}$, except for the fields 
$F_0$ and $C_1^I$, which as we will explain later constitute initial data, 
whose evolutions are determined by the conservation equations.  
The evolution equations then determine the evolution of 
$D_{IJ}$ and $E_{IJ}$ (as well as the other terms at higher powers
in the $1/r$ expansion of $h_{IJ}$), given prescribed initial data.  
Rather unusually, $C_{IJ}(u,x^I)$ constitutes free data.  Furthermore, 
note that the initial data are unconstrained, in contrast to a 
Cauchy formulation where they would satisfy elliptic constraints.  
The explicit Einstein equations for the metric components are listed in section 2.2 of Ref.\ \cite{fakenews}.

In summary, the Einstein equation is solved by prescribing initial 
data $$\{F_0(u_0, x^I), C_1^I(u_0, x^I), D_{IJ}(u_0, x^I), E_{IJ}(u_0, x^I), 
\ldots\},$$ where the ellipses denote terms at higher powers 
in the $1/r$ expansion of $h_{IJ}$, and an arbitrary trace-free 
symmetric tensor $C_{IJ}(u,x^I).$  The hypersurface equations then determine 
all scalar, vector and tensor fields in the expansions 
(\ref{met:falloff}) on the initial hypersurface.  Now, the evolution and 
conservation equations can be integrated to determine 
$\{F_0, C_1^I, D_{IJ}, E_{IJ}, \ldots\}$ at a time step $u_0 + \Delta u,$ 
before iterating the above procedure to find the full solution at this time 
step and so on. 

One of the hypersurface equations, arising at order 
$1/r^3$ of the $\ell^\mu\, m^\nu\, G_{\mu\nu}=0$ equation 
(see equation (2.16) of Ref.\ \cite{fakenews}), 
gives\footnote{$I,J, \ldots$ 
indices are raised or lowered using $\omega^{IJ}$ or $\omega_{IJ}$.} 
\be
C_0{}^I = -\ft12 D_J\, C^{IJ}\,,\label{CIJeqn}
\ee
which determines $C_0^I$ given the data $C_{IJ}(u,x^I)$.  Alternatively, 
we may view this equation as determining $C_{IJ}$ given a choice of 
$C_0^I$, up to functions of integration which may themselves be arbitrarily
chosen.  That is to say, we may exchange the freedom to choose a 
trace-free tensor $C_{IJ}(u,x^I)$ with the freedom to choose a 
vector $C_0^I$.  It will be useful in what follows to take this 
perspective.  Therefore, the data for the characteristic value 
problem are given by
$$\{F_0(u_0, x^I), C_0^I(u,x^I),\, C_1^I(u_0, x^I), D_{IJ}(u_0, x^I), E_{IJ}(u_0, x^I), \ldots\},$$
where we emphasise that the choice for $C_0^I$ is not only an initial value 
choice as with the other fields, but that we have the freedom to 
choose the vector completely, as a function also of $u$.

Let us consider, briefly, the physical significance of these quantities.  
The Bondi mass and angular momentum are given by 
\cite{bondi, sachs}, \cite{softnuts}\footnote{We work in units where $G=1$. Note
that the total derivative term $D_I D_J C^{IJ}$ in the expression 
for the Bondi mass arises
because the tensor $C^{IJ}$ is not necessarily regular on the sphere
\cite{softnuts}.}
\begin{equation} \label{Bondi:mj}
 M_B = - \frac{1}{8\pi} \int_S d\Omega\, \Big(F_0 +
 \fft14 D_I D_J C^{IJ}\Big), \qquad J_B = \frac{3}{8\pi} 
\int_S d\Omega\, \sin^2 \theta\, L^\phi,
\end{equation}
where 
\begin{equation} \label{angmtm}
 L^I = \frac{1}{2} C_1^I + \frac{1}{3} C^{IJ} C_{0\, J}
\end{equation}
is the angular momentum aspect, so that if $L^\phi$ is constant then
$J_B = L^\phi$.  Moreover, we have that the dual Bondi mass or Bondi NUT charge is given by \cite{softnuts}
\begin{equation}
 \widetilde{M}_B = \frac{1}{16\pi} \int_S d C_0,  \label{NUT}
\end{equation}
where $C_0=C_{0\, I}\, dx^I.$
Thus $F_0$, $C_0^I$ and $C_1^I$ constitute the data determining
the Bondi mass, NUT charge and angular momentum.  
Furthermore, the other data given by terms at higher powers in the $1/r$ 
expansion of $h_{IJ}$ correspond to subleading BMS charges 
\cite{fakenews}, \cite{dualex}.  For example, $D_{IJ}$ determines the BMS 
charge at order $1/r^2.$

In this paper, our focus will be on constructing stationary solutions, meaning that all the scalar,
vector and tensor fields in the expansions (\ref{met:falloff}) will
be independent of the retarded time $u$.   This means that the
evolution  and conservation equations (\ref{uderivs}) and \eqref{cons} will now give rise to genuine
constraints on the initial data.  Thus, all stationary asymptotically flat solutions are given by prescribing the following data
$$\{F_0(x^I),\, C_0^I(x^I),\, C_1^I(x^I),\, D_{IJ}(x^I),\, E_{IJ}(x^I), \ldots\} $$
subject to the constraints implied by equations 
\eqref{uderivs} and \eqref{cons}.

\section{$C_{0\, I}$ as a Maxwell gauge field}

  As can be verified from the
form of the BMS transformations in the notation of Ref.\ \cite{fakenews} that
we are using here, under a supertranslation with diffeomorphism
parameter $\xi= s(\theta,\phi)\, \del u+ \cdots$, we have
\begin{equation}
 \delta  C_{0\, I} = \partial_u C_0^I +  D_I \left(\frac{1}{2}\Box s + s \right).
\end{equation}
Thus, if 
\begin{equation}
 \partial_u C_0^I = 0,
\end{equation}
which is the case for stationary solutions, but also more generally, then
\begin{equation} \label{C0gauge}
{} \hspace{10mm} \delta  C_{0\, I} = D_I \Lambda,  \qquad  \qquad \Lambda = \frac{1}{2}\Box s + s
\end{equation}
Thus, in a very real sense, one can think of a $u$-independent 1-form $C_0$
as being analogous to a Maxwell gauge potential.

This interesting observation is one reason why we chose to view $C_{0\, I}$ rather than $C_{IJ}$ as characteristic data in the previous section.

\section{Dirac monopole}

Given the interpretation of a $u$-independent $C_0$ as a Maxwell gauge field, it is natural to consider the case where this 1-form describes a Dirac magnetic
monopole on the 2-sphere, with $C_0$ of the form
\be
C_0\equiv C_{0\, I}\, dx^I = 2 p \cos\theta\, d\phi,\label{C0ans0}
\ee
where $p$ is a constant. This 1-form is singular at the north and the
south poles of the sphere.  We shall look for solutions that are
stationary and axisymmetric,\footnote{The most general such solutions 
have been classified in Weyl coordinates, and are given by
four functions that satisfy simple coupled partial differential 
equations in two variables.  For details, see chapter 20 of
Ref.~\cite{stephani}.} so all the metric functions will be 
assumed to be independent of $u$ and of $\phi$.  Of course, in order 
to specify a particular solution we must also prescribe initial data for 
the other fields.  However, since we will have constraint equations 
implied by the requirement of stationarity, we choose for now to keep the 
other initial data arbitrary and choose them with the constraint 
equations in mind.

   The singularities
in the Dirac 
monopole gauge potential $A=2p\cos\theta\, d\phi$ are purely gauge
artefacts, with the field strength $F=dA=-2p \sin\theta\, d\theta\wedge d\phi$
being perfectly regular on the sphere.  The Dirac string or wire
singularities in the gauge potential can be moved around by means
of large gauge transformations $A\rightarrow A + d\Lambda$.  For example,
taking $\Lambda=-2p\, \phi$ gives $A=-4p \sin^2\fft{\theta}{2}\, d\phi$,
which is singular only at the south pole, whilst taking $\Lambda=2p\, \phi$
gives $A=4p\cos^2\fft{\theta}{2}\, d\phi$, which is singular only
at the north pole.  

  By comparing with equation \eqref{C0gauge}, and taking $s$ to be a 
constant multiple of the
azimuthal coordinate $\phi$, we can perform precisely the kinds of 
gauge transformation we described above for the  Dirac
monopole.\footnote{The function $\phi$ is of course singular on the
sphere.  However, since we are already entertaining the idea of
using a monopole configuration for $C_{0\, I}$ that is singular
on the sphere, there is no longer any reason to restrict ourselves
to non-singular supertranslation parameters.}
We may take as our starting point the slight
generalisation of (\ref{C0ans0}) where
\be
C_0\equiv C_{0\, I}\, dx^I = 2 p \cos\theta\, d\phi+ 2k\, d\phi,\label{C0ans}
\ee
with $k$ being a gauge-adjustable constant, which fixes the 
supertranslation gauge that we are working in.  
Already, it is clear that we are dealing with spacetimes that have
a non-trivial NUT charge \eqref{NUT}:
\begin{equation}
 \widetilde{M}_B = -\frac{p}{2}.
\end{equation}

   Using the fact that $C_{IJ}$ is symmetric, trace-free and depends 
only on $\theta$, by the axisymmetry assumption, we can 
substitute (\ref{C0ans}) into (\ref{CIJeqn}) and
solve to find
\be 
C_{\theta\theta}=\fft{c_1}{\sin^2\theta}\,,\qquad C_{\phi\phi}= -c_1\,,\qquad
C_{\theta\phi}= \fft{c_2 + 4 k \cos\theta +2 p \cos^2\theta}{\sin\theta}\,,
\label{CIJsol}
\ee
where $c_1$ and $c_2$ are constants of integration.  These constants
will contribute to subleading BMS charges, and as is evident from
the definition (\ref{angmtm}) of the angular momentum aspect $L^I$ and the
form of (\ref{C0ans}), $c_1$ will also contribute to the angular momentum.
Note that $D_I D_J C^{IJ}=0$, and therefore $c_1$ and $c_2$ do not 
contribute to the Bondi mass given in \eqref{Bondi:mj}.

It is interesting to
note that although in general this solution for $C_{IJ}$ is singular
at one or both of the poles of the sphere, we can find a non-singular solution
in the special case where we choose $c_1=0$, $c_2=-2p$ and  $k=0$, for which
the only non-vanishing components of the symmetric $C_{IJ}$ are 
specified just by
\be 
C_{\theta\phi}=-2p \sin\theta.
\ee
Of course, since our starting point was $C_{0\, I}$ having the form
of a Dirac monopole, which is necessarily singular somewhere on the sphere,
there is no particular reason why we should expect or require $C_{IJ}$ to 
be non-singular.  Indeed, as we shall see below, there are reasons to
prefer different assignments for $(c_1,c_2,k)$ for which $C_{IJ}$
does have singularities.

As explained before, the other hypersurface equations are algebraic equations that determine the form of other fields given initial data.  Therefore, we need not concern ourselves with those just yet.  However, the evolution and conservation equations are now non-trivial equations, constraining the data.  Therefore, we focus on these equations.
  
The $n^\mu\, n^\nu\, G_{\mu\nu}=0$
projection of the 
Einstein equation at order $1/r^2$ implies the conservation equation 
\cite{fakenews}
\be
\del_u F_0= -\frac{1}{2} D_I D_J \partial_u C^{IJ} 
+ \frac{1}{4} \partial_u C^{IJ} \partial_u C_{IJ}, \label{uF0} 
\ee
which is trivially satisfied in the stationary case.  The other conservation 
equation, which comes from the $n^\mu\, m^\nu\, G_{\mu\nu}=0$ projection 
of the Einstein equation at order $1/r^3$, then gives (see equation 
(2.26) of \cite{fakenews}) 
\bea
0 = 3 \partial_u C_1^I &=& D^I F_0 - D_J (D^J C_0^{I} - D^I C_0^{J}) \nn \\
&=& D^I F_0 - D_J F^{JI}, \label{uC1}
\eea
where $F_{IJ} = 2 \partial_{[I} C_{0\, J]}$ is the field strength associated 
with the gauge field $C_0$. In the first line we have used equation 
\eqref{CIJeqn}, and the fact that $\partial_u C_{IJ} = 0$.  Since the 
only non-zero component of $F^{IJ}$ is 
$F^{\theta \phi} = -2p/\sin \theta$, and 
\begin{equation}
 D_J F^{JI} = \frac{1}{\sin \theta} \partial_J (\sin \theta F^{JI})=0,
\end{equation}
equation \eqref{uC1} then implies that $F_0$ is independent of the 
coordinates on the sphere. Hence, we choose
\be
F_0=-2m,
\ee
with constant $m$ parameterising the mass of the asymptotically-flat spacetime.  

   Moving on to the evolution equations given by the $m^\mu\, m^\nu\, G_{\mu\nu}=0$ projection of the 
Einstein equation, at
order $1/r^4$, this gives  \cite{fakenews}
\bea
0 = \partial_u D_{IJ} &=& - \frac{1}{4} F_{0} C_{IJ}  - \frac{1}{2} D_{(I} C_{1\, J)} + \frac{1}{4} C_{IJ} D_K C_0^K + \frac{1}{32} D_I D_J C^2 \nn\\
&&- D_{(I}(C_{J)K} C_0^{K}) - \frac{1}{8} D_I C^{KL} D_J C_{KL}    \\
&&+ \frac{1}{4} \omega_{IJ} \Big[ D_K C_1^K -\frac{1}{1
6} \Box C^2 + 2 D^{K}(C_{KL} C_0^{L}) +  \frac{1}{4} D^M C^{KL}  
D_{M} C_{KL} \Big]. \nn \label{uD}
\eea
With $C_0$ and $C_{IJ}$ given by \eqref{C0ans} and (\ref{CIJsol}), respectively, we may now seek to 
solve for $C_{1\, I} (\theta)$.
Since the solutions
are a little complicated in general we shall not present them here, but just
remark that in general they involve terms that have power-law singularities at
$\theta=0$ and $\theta=\pi$, and also terms proportional to 
$\log\cot\fft{\theta}{2}$ with logarithmic singularities at the poles of the
sphere.  The solution for $C_{1\, I}$ will be free of logarithmic
singularities if and only if we choose the $c_1$ and $c_2$ integration 
constants in (\ref{CIJsol}) to be 
\be
c_1=0\,,\qquad c_2= 2p.\label{c1c2choice}
\ee
The solution is then given by
\bea
C_{1\, \theta}&=& c_3\, \sin\theta + \fft{3k \cos4\theta +
  4(k^2+p^2)\cos3\theta -12 k p \cos2\theta -36(k^2+p^2) \cos\theta -
   55 kp}{4\sin^5\theta}\,,\nn\\
C_{1\, \phi}&=& c_4 \sin^2\theta - 4m (p\cos\theta +k),\label{C1Isol}
\eea
where $c_3$ and $c_4$ are constants of integration.

This is the most general solution for $C_{1\, I}$ without logarithmic 
singularities, starting from the Dirac monopole connection (\ref{C0ans}). 
We may now evaluate the angular momentum
given in \eqref{Bondi:mj}, obtaining
\be
J_B=  \frac{1}{2}(c_4 - 6 k m).
\ee
However, the Komar angular momentum
\begin{equation}
 J_K = \frac{1}{16 \pi}  \int_S \star d j^\flat 
\end{equation}
with $j=\partial/\partial \phi$ is divergent.  This is easy to see, since
\begin{equation}
 (\star d j^\flat)_{\theta \phi} = 2[2k \sin\theta + p \sin 2\theta]\, r + [2(c_4 - 6km) \sin\theta - 6 p m \sin2\theta] + {\cal O}(r^{-1}).
\end{equation}
Therefore, before sending $r$ to infinity, the
right hand side of the expression for $J_K$ gives
\begin{equation}
kr+ \frac{1}{2}(c_4 - 6 k m) + {\cal O}(r^{-1}).
\end{equation}
A simple explanation for why the divergence above vanishes for $k=0$ is that 
the wire singularity provides a divergent contribution, however, when $k=0$ 
the divergent contributions from the wire singularities at the north and 
south poles cancel.  
There has been a proposal to resolve this divergence issue for general $k$ 
in  Ref.\ \cite{Bordo:2019rhu}, by redefining the Komar integral.  
However, given that $k$ parameterises a large gauge transformation, 
we may simply avoid these difficulties by choosing to set $k=0$. 
(Recall that $k$ is the gauge parameter appearing in \eqref{C0ans}.)  Now,
we obtain
\be
J_B= \fft{1}{2} c_4.\label{anm}
\ee
Thus, $c_4$ is a Kerr-like angular momentum parameter, which, for simplicity,
we shall for now set to zero. The integration constant $c_3$ will contribute to
subleading BMS charges \cite{fakenews}, \cite{dualex}, and we shall also set this to zero for simplicity.

  With $c_3=c_4=0$, and making the gauge
choice $k=0$ as discussed above, the 
expressions for $C_{1\, I}$ in (\ref{C1Isol}) become
\be
C_{1\, \theta}= -\fft{4p^2\, (2+\sin^2\theta)\cos\theta}{\sin^5\theta}\,,
\qquad C_{1\, \phi}= -4 m p\, \cos\theta.
\ee
The previous expressions (\ref{C0ans}) and (\ref{CIJsol}) for
$C_{0\, I}$ and $C_{IJ}$ then  give
\bea
 C_{0\, \theta}&=&0\,,\qquad C_{0\, \phi} = 2p\cos\theta,\nn\\
C_{\theta\theta}&=&0\,,\qquad C_{\phi\phi}=0\,,\qquad 
C_{\theta\phi}= \fft{2p\, (1+\cos^2\theta)}{\sin\theta}.
\eea
Substituting these expressions into the $\ell^\mu\, n^\nu\, G_{\mu\nu}=0$
component of the Einstein equation at order $1/r^4$ then gives
\be
F_1= \fft{p^2\, (4+4\sin^2\theta - 11 \sin^4\theta)}{2\sin^4\theta}.
\ee
Comparing with the expressions in appendix A,
we see that the above fields are in exact agreement with the
corresponding components in the expansion of the Taub-NUT metric
in Bondi coordinates, with $p=\ell$.  

  Clearly we could in principle continue to arbitrarily high 
orders in the $1/r$ expansions 
of the Einstein equation.  In particular, the structure of the constraint 
equations coming from the evolution equations should be clear now.  For
example, at 
the next order (see equation (2.21) of \cite{fakenews}), the equation implied by the $u$-independence of $E_{IJ}$ implies an ordinary differential equation for $D_{IJ}(\theta)$.  
Following this iterative process, and making appropriate choices for
constants of integration as they arise, 
reproduces the Taub-NUT metric in Bondi coordinates to any desired 
order.  Note that had we not chosen to set $k=0$, we could have 
obtained the Bondi metric corresponding to the Taub-NUT metric with
the string singularity correspondingly shifted.

   Other choices of the integration constants will give
more general solutions, generically presumably 
with more severe singular behaviour on the sphere, such as the logarithmic
behaviour in $C_{1\, I}$ that we avoided above by the judicious choice
(\ref{c1c2choice}) for the constants of integration $c_1$ and $c_2$. 

  We have already seen, \eqref{anm}, that the constant of
 integration $c_4$ in the
expression (\ref{C1Isol}) for $C_{1\, \phi}$ is related to angular momentum. 
In fact, if we turn off the Dirac monopole altogether by 
choosing $p=0$ (and $k=0$), i.e.~setting $C_{0\, I}=0$, and 
letting $c_3=0$ and $c_4= 2 m a$, we find that our solution 
\eqref{C1Isol} becomes
\be
C_{1\, \theta}=0\,,\qquad C_{1\, \phi}= 2 m a \sin^2\theta.\label{kerrC1}
\ee
As 
can be seen from
    the expressions in appendix A of \cite{softnuts}, 
the $C_{1\,I}$ above is precisely that of the Kerr metric
in Bondi coordinates, 
where $a$ is the Kerr rotation parameter.
In fact the Kerr metric could be derived in Bondi coordinates by using
the same iterative technique we have illustrated in this paper for the 
Taub-NUT metric, by starting with $F_0=-2m$, $C_{0\, I}=0$ 
and $C_{1\, I}$ given by
(\ref{kerrC1}), and then sequentially solving the Einstein equation order
by order in powers of $1/r$ with the evolution equations 
determining lower order powers of $h_{IJ}$, under the assumption of 
stationarity. Of course one could also derive the 
Kerr-Taub-NUT metric in Bondi coordinates, by starting out as we did above 
for Taub-NUT, but now at the stage where we obtained the expressions 
(\ref{C1Isol}) for $C_{1\, I}$ we would take $c_4$ to be non-zero.

\section{Discussion}

We showed that, in a stationary setting, starting from a choice of the 
gauge field $C_0$ as the Dirac monopole, we can integrate the 
vacuum Einstein equation to recover the Taub-NUT solution in Bondi 
coordinates.  Assuming the solution to be stationary, the evolution and 
conservation equations transformed into constraint equations on the 
other characteristic data.  For example, we found that $F_0$ must be 
constant, giving a Bondi mass.  Solving another of the constraint equations, 
we found $C_1^I$ up to two constants of integration.  One of these 
constants corresponded to a Kerr-like angular momentum parameter.  
Working through the other constraint equations would also generate other 
constants, which contribute to subleading BMS charges.  Therefore, 
what the Dirac monopole actually 
generates is a \emph{family} of Taub-NUT-like solutions, of which the 
usual Taub-NUT solution is one member.  In general, the other members
of the family would correspond to other stationary axisymmetric Weyl
solutions with a non-trivial NUT charge.

The 1-form  $C_0$ can in fact be regarded as a gauge connection
more generally, namely in any spacetime, even a time-dependent one,
provided that $C_0$ itself is still time-independent. 
Therefore the iterative procedure that we carried out, starting with
a Dirac monopole configuration, can be repeated in a 
more general non-stationary setting. It would be interesting to see 
if dynamical solutions with Taub-NUT charge can be constructed this way. 

Moreover, in this work, for simplicity, we have assumed a $1/r$-expansion 
of the Bondi metric, at least for the first few orders.  However, this is 
not necessary and only simplifies the hypersurface, \eqref{constraints}, 
and evolution, \eqref{uderivs}, equations. Therefore, one could consider the more general 
system of equations with weaker fall-off conditions, which would 
presumably be necessary when considering time-dependent solutions.

\section*{Acknowledgements}

M.G.\, and C.N.P.\, would like to thank the Max-Planck-Institut f\"ur Gravitationsphysik (Albert-Einstein-Institut), Potsdam where this work was initiated.  M.G.\ is supported by a Royal Society University Research Fellowship. C.N.P.\ is partially supported by DOE grant DE-FG02-13ER42020.

\appendix

\section{Taub-NUT Metric in Bondi Coordinates}

   Here, we construct the Taub-NUT metric in Bondi coordinates,
working in the ``symmetric'' coordinate gauge where there are string
singularities at both the north and south poles.   The starting metric 
in this case is the standard one,
\be
ds^2 = -f(\bar r)\,(d\bar t +2\ell\,\cos\bar\theta\,d\bar\phi)^2 
+f(\bar r)^{-1}\,d\bar r^2+
   (\bar r^2\ell^2)\,(d\bar\theta^2+\sin^2\, \bar\theta\,d\bar\phi^2)\,,
\label{TNmetsouth}
\ee
where $f(\bar r) = (\bar r^2 -2 m \bar r -\ell^2)(\bar r^2+\ell^2)^{-1}$. We have placed bars on the coordinates, because we now make an expansion
of the form described in \cite{softnuts}, imposing the Bondi metric conditions 
$g_{rr}=g_{u\theta}=g_{u\phi}=0$, and $\det(h_{IJ})=\det(\omega_{IJ})$, 
order by order in the expansion in $1/r$.  Proceeding to
the first few orders, we find 
{\setlength{\jot}{10pt}
\crampest
\bea
\bar t&=& u+r + 2m \log r -\fft{4m^2 -2 \ell^2\, (\csc^2\theta+\csc^4\theta
-\fft{11}{4})}{r} 
- \fft{2 m\,[2 m^2 + \ell^2\, (3-2\csc^2\theta)] }{r^2} 
+\cdots\,,\nn\\
\bar\phi&=& \phi -\fft{2\ell \cos\theta}{r\, \sin^2 \theta} -
\fft{\ell^3\, (\sin^4 \theta + 4 \sin^2 \theta - \fft{20}{3})}{
      r^3\, \sin^6 \theta}  -\fft{2m\,\ell^3 \,\cos^3\theta}{
                        r^4\, \sin^4 \theta} +\cdots\,,\nn\\
\bar r &=& r + \fft{\ell^2\,(4\csc^4\theta-3)}{2r} 
- \fft{2m\ell^2\,\cot^2\theta}{r^2} \nn \\
&&
+\fft{\ell^4\,(-6 +16 \sin^2 \theta + 11 \sin^4 \theta - 25 \sin^6 \theta + \fft{45}{8} \sin^8 \theta)}{3 r^3 \, \sin^8\theta} +
\fft{m\ell^4 \,(7\cos2\theta+1)\cos^2\theta}{r^4\,\sin^6\theta}+\cdots\,,
\nn\\
\bar\theta &=& \theta  +\fft{2\ell^2\,\cos\theta}{r^2\,\sin^3\theta} +
\fft{2 \ell^4\,(\sin^4 \theta + 10 \sin^2 \theta -15)\, \cos \theta}{3r^4 \, \sin^7 \theta} +
\fft{4m\ell^4\,\cos^3\theta}{r^5\, \sin^5\theta}+\cdots\,.
\eea
\uncramp
}
We have actually worked to a higher order than the terms presented here,
sufficient for our later purposes.  
Using these expansions, we then obtain the Taub-NUT metric in Bondi form,
finding
{\setlength{\jot}{10pt}
\bea
g_{uu} &=& -1 +\fft{2m}{r} + \fft{2\ell^2}{r^2} +
\fft{m\ell^2 \, (1-4\csc^4\theta)}{r^3}
+ \fft{4 \ell^4\, (\sin^4 \theta - 2) + m^2 \ell^2 \,\sin^2 2\theta}{r^4\, \sin^4 \theta}
  +{\cal O}(r^{-5})\,,\nn\\
g_{ur} &=& -1 +\fft{\ell^2\,(1+\cos^2\theta)^2}{2 r^2\,\sin^4\theta}-
\fft{\ell^4\,[6 - 28 \sin^2 \theta +15 \sin^4 \theta +8 \sin^6 \theta -\fft{21}{8} \sin^8 \theta]} {r^4\,\sin^8\theta} 
+{\cal O}(r^{-5})\,,\nn\\
g_{u\theta}&=& \fft{4 \ell^2\,(2 - \sin^2 \theta + \sin^4 \theta)\,\cos\theta}{r\, \sin^5 \theta} +\fft{4 m\ell^2\, (\cot^2\theta-1) \cot \theta}{r^2} \nn\\
&&-
\fft{2\ell^4\,(8 - 28 \sin^2 \theta +\fft{50}{3} \sin^4\theta -4 \sin^6 \theta +3 \sin^8 \theta)\,\cos\theta}{r^3\, \sin^9 \theta}+ 
   {\cal O}(r^{-4})\,,\nn\\
g_{u\phi} &=& -2\ell\,\cos\theta + 
\fft{4m\ell \cos\theta}{r}+\fft{4\ell^3\,\cos\theta}{r^2}
 +
\fft{2 m \ell^3\,(1-4\csc^4\theta)\cos\theta}{r^3} 
    + {\cal O}(r^{-4})\,,\nn\\
g_{rr}&=& {\cal O}(r^{-6})\,,\qquad g_{r\theta}={\cal O}(r^{-5})\,,\qquad
g_{r\phi}= {\cal O}(r^{-5})\,,\nn\\
g_{\theta\theta}&=& r^2 + \fft{2 \ell^2\, (1+\cos^2\theta)^2}{\sin^4\theta} -
 \fft{4m\ell^2\,\cot^2\theta}{r} \nn\\
&&+
 \fft{2\ell^4\,(40 - 16 \sin^2 \theta - 45 \sin^4 \theta +27 \sin^6 \theta)}{
  3 r^2\,\sin^6\theta}  + {\cal O}(r^{-3})\,,\nn\\
g_{\theta\phi}&=& \fft{2\ell r\,(1+\cos^2\theta)}{\sin\theta} +
  \fft{\ell^3\, (8 -\fft{20}{3} \sin^2 \theta + 6 \sin^4 \theta - 5 \sin^6 \theta)}{r\,\sin^5\theta} 
 -\fft{10m \ell^3\,\cos^2\theta}{r^2\,\sin\theta} 
  + {\cal O}(r^{-3})\,,\nn\\
g_{\phi\phi}&=& r^2\sin^2\theta +
\fft{2\ell^2\, (1+\cos^2\theta)^2}{\sin^2\theta} 
+ 
\fft{4 m\ell^2\, \cos^2\theta}{r}\nn\\
&&+
\fft{2\ell^4\,(8 - \fft{16}{3} \sin^2 \theta - \sin^4\theta -\sin^6\theta)}{r^2\,\sin^4\theta} 
 + {\cal O}(r^{-3})\,.
\eea
}
Comparing with the expansions for the Bondi metric as defined in
(\ref{met:falloff}), we have, for example,
\bea
C_{IJ}:&& C_{\theta\theta}=0\,,\qquad C_{\phi\phi}=0\,,\qquad
C_{\theta\phi}= 2 \ell\,(1+\cos^2\theta)\, \csc\theta\,,\nn\\
D_{IJ}: && D_{\theta\theta}=-4m\ell^2 \,\cot^2\theta\,,\qquad
D_{\phi\phi}=4m\ell^2 \,\cos^2\theta\,,\nn\\
&& D_{\theta\phi}= \ell^3\, (8 -\textstyle{\fft{20}{3}} \sin^2 \theta + 6 \sin^4 \theta - 5 \sin^6 \theta)\,\csc^5\theta \,,\nn\\
C_{0\,I}:&& C_{0\, \theta}=0\,,\qquad 
C_{0\, \phi} =  2\ell \cos\theta\label{C0CIJ}\,,\nn\\
C_{1\, I}:&&  C_{1\,\theta}= 
  -4\ell^2\,(2+ \sin^2\theta)\cot\theta\csc^4\theta\,,\qquad
C_{1\,\phi}= -4m\ell\,\cos\theta\,,\nn\\
F_0&=& -2m\,,\qquad F_1=\fft{\ell^2\,(4+ 4 \sin^2\theta-11 \sin^4\theta)}{
            2\sin^4\theta}\,.
\eea

\bibliographystyle{utphys}
\bibliography{NP}

\providecommand{\href}[2]{#2}\begingroup\raggedright\begin{thebibliography}{10}

\bibitem{Bishop2016}
N.~T. Bishop and L.~Rezzolla,
  \href{http://dx.doi.org/10.1007/s41114-016-0001-9}{``Extraction of
  gravitational waves in numerical relativity,''{\em Living Reviews in
  Relativity} {\bf 19} (Oct, 2016)  2}.

\bibitem{Strom:lec}
A.~Strominger, ``{Lectures on the Infrared Structure of Gravity and Gauge
  Theory},''
\href{http://arxiv.org/abs/1703.05448}{{\tt arXiv:1703.05448 [hep-th]}}.

\bibitem{Compere:2018aar}
G.~Comp\`ere and A.~Fiorucci, ``{Advanced Lectures on General Relativity},''
\href{http://arxiv.org/abs/1801.07064}{{\tt arXiv:1801.07064 [hep-th]}}.

\bibitem{dual0}
H.~Godazgar, M.~Godazgar, and C.~N. Pope, ``{New dual gravitational charges},''
  \href{http://dx.doi.org/10.1103/PhysRevD.99.024013}{{\em Phys. Rev.} {\bf
  D99} (2019) no.~2, 024013},
\href{http://arxiv.org/abs/1812.01641}{{\tt arXiv:1812.01641 [hep-th]}}.

\bibitem{dualex}
H.~Godazgar, M.~Godazgar, and C.~N. Pope, ``{Tower of subleading dual BMS
  charges},'' \href{http://dx.doi.org/10.1007/JHEP03(2019)057}{{\em JHEP} {\bf
  03} (2019)  057},
\href{http://arxiv.org/abs/1812.06935}{{\tt arXiv:1812.06935 [hep-th]}}.

\bibitem{porrati}
U.~Kol and M.~Porrati, ``{Properties of Dual Supertranslation Charges in
  Asymptotically Flat Spacetimes},''
\href{http://arxiv.org/abs/1907.00990}{{\tt arXiv:1907.00990 [hep-th]}}.

\bibitem{softnuts}
H.~Godazgar, M.~Godazgar, and C.~N. Pope, ``{Dual gravitational charges and
  soft theorems},''
\href{http://arxiv.org/abs/1908.01164}{{\tt arXiv:1908.01164 [hep-th]}}.

\bibitem{Winicour2009}
J.~Winicour, \href{http://dx.doi.org/10.12942/lrr-2009-3}{``Characteristic
  evolution and matching,''{\em {Living Reviews in Relativity}} {\bf 12} (Apr,
  2009)  3}.

\bibitem{Monteiro:2014cda}
R.~Monteiro, D.~O'Connell, and C.~D. White, ``{Black holes and the double
  copy},'' \href{http://dx.doi.org/10.1007/JHEP12(2014)056}{{\em JHEP} {\bf 12}
  (2014)  056},
\href{http://arxiv.org/abs/1410.0239}{{\tt arXiv:1410.0239 [hep-th]}}.

\bibitem{Luna:2015paa}
A.~Luna, R.~Monteiro, D.~O'Connell, and C.~D. White, ``{The classical double
  copy for Taub-NUT spacetime},''
  \href{http://dx.doi.org/10.1016/j.physletb.2015.09.021}{{\em Phys. Lett.}
  {\bf B750} (2015)  272--277},
\href{http://arxiv.org/abs/1507.01869}{{\tt arXiv:1507.01869 [hep-th]}}.

\bibitem{fakenews}
H.~Godazgar, M.~Godazgar, and C.~N. Pope, ``{Subleading BMS charges and fake
  news near null infinity},''
  \href{http://dx.doi.org/10.1007/JHEP01(2019)143}{{\em JHEP} {\bf 01} (2019)
  143},
\href{http://arxiv.org/abs/1809.09076}{{\tt arXiv:1809.09076 [hep-th]}}.

\bibitem{sachs}
R.~K. Sachs, ``{Gravitational waves in general relativity: 8. Waves in
  asymptotically flat space-times},''
\href{http://dx.doi.org/10.1098/rspa.1962.0206}{{\em Proc. Roy. Soc. Lond.}
  {\bf A270} (1962)  103--126}.

\bibitem{bondi}
H.~Bondi, M.~G.~J. van~der Burg, and A.~W.~K. Metzner, ``{Gravitational waves
  in general relativity: 7. Waves from axisymmetric isolated systems},''
\href{http://dx.doi.org/10.1098/rspa.1962.0161}{{\em Proc. Roy. Soc. Lond.}
  {\bf A269} (1962)  21--52}.

\bibitem{stephani}
H.~Stephani, D.~Kramer, M.~A.~H. MacCallum, C.~Hoenselaers, and E.~Herlt,
  \href{http://dx.doi.org/10.1017/CBO9780511535185}{{\em {Exact solutions of
  Einstein's field equations}}}.
\newblock Cambridge Monographs on Mathematical Physics. Cambridge Univ. Press,
  Cambridge,
2003.
\newblock

\bibitem{Bordo:2019rhu}
A.~B. Bordo, F.~Gray, R.~A. Hennigar, and D.~Kubiznak, ``{The First Law for
  Rotating NUTs},''
\href{http://arxiv.org/abs/1905.06350}{{\tt arXiv:1905.06350 [hep-th]}}.

\end{thebibliography}\endgroup

\end{document}